\documentclass[aps,prl,twocolumn,showpacs]{revtex4}
\usepackage{latexsym,graphicx,eufrak}
\newcommand{\draftfig}[1]{\includegraphics[angle=0,scale=0.4]{#1}}
\newcommand{\draftfigb}[1]{\includegraphics[angle=0,scale=0.32]{#1}}

\begin{document}
\title{A Theory of Inertial Range Similarity in Isotropic Turbulence}
\author{Mogens V. Melander and Bruce R. Fabijonas}
\affiliation{
Department of Mathematics,
Southern Methodist University,
Dallas, TX 75275}
\date{\today}
\begin{abstract}
We consider equilibrium statistics for high Reynolds number isotropic 
turbulence in an incompressible flow driven by steady forcing at the largest 
scale. Motivated by shell model observations, we develop a similarity theory 
for the inertial range from clearly stated assumptions. In the right 
variables, the theory is scaling invariant, but in traditional variables it 
shows anomalous scaling. We obtain the underlying probability density 
function, the scaling exponents, and the coefficients for the structure 
functions. An inertial range length scale also emerges.
\end{abstract}
\maketitle

\begin{figure}[b]
\begin{center}
\draftfig{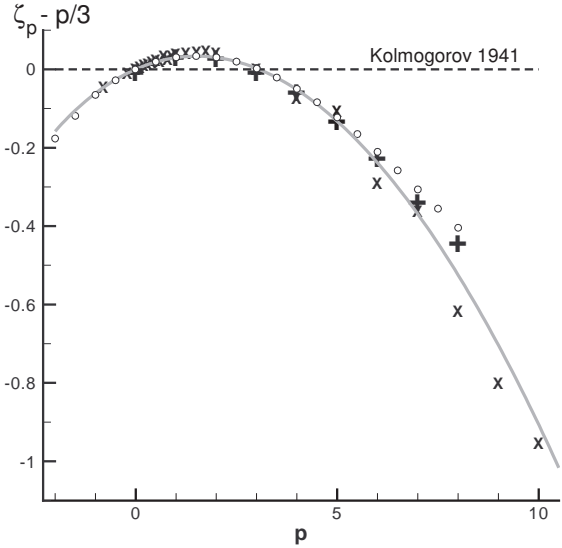}
\caption{Compensated exponents. Experimental data: longitudinal 
(+) from \cite{benzi:cili:baudet:chav:95}, transverse (x) from \cite{kurien:srini:01}. 
Our shell model data (o); note the 
statistical convergence is poor for $p>5$. The curve is our theory with 
$\beta =1.83$ and $\zeta _{-2} =-0.82$.
\label{fig:one}
}
\end{center} 
\end{figure}
\begin{figure}[t]
\begin{center}
\draftfig{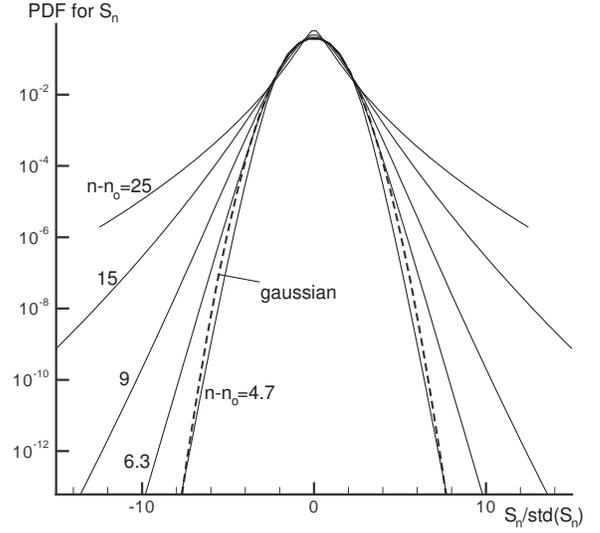}
\caption{Normalized one dimensional PDF obtained from our theory. 
$n$ is the shell number which increases towards smaller scales.
\label{fig:two}
}
\end{center}
\end{figure}
Turbulence is often called the last great unsolved problem in classical 
physics. Being responsible for rapid mixing and energy dissipation in 
fluids, turbulence is also critical in many applications. 
Turbulence research aims to describe an ensemble of flow realizations 
statistically using moments or probability density functions (PDFs). The 
simplest and most symmetric problem to consider is steady-forced isotropic 
turbulence in an incompressible fluid. 
At high Reynolds numbers, large-scale forcing is 
separated from the dissipation by a wide range of scales where inertial 
forces rule and the physics is universal. In 1941, Kolmogorov~\cite{K41} developed the 
first theory for this inertial range. His theory removes scale dependence by 
normalization. Unfortunately, his scale invariance is at odds with 
observations. 
The discrepancy is called ``anomalous scaling." According to 
Frisch \cite[p. ix]{frisch}, ``modern 
work on turbulence focuses to a large extent on 
trying to understand the reasons for the partial failure of the 1941 
theory." 

In the inertial range, the $p$th order moment or structure function $\mathcal{S}_p $ for 
the velocity difference $\delta v(l)$ 
between two points is a power law in the separation 
distance $l$ \cite{frisch}, i.e., $\mathcal{S}_p \propto l^{\zeta _p }$. 
Kolmogorov \cite{K41,frisch}
suggested $\zeta _p =p/3$. This way, $\mathcal{S}_p (l)/\mathcal{S}_2^{p/2} (l)$ 
is independent 
of $l$. Moreover, the PDFs for $\delta v$ at 
different $l$ collapse to a single curve when 
plotted in units of standard deviations. For a wide inertial range, 
Kolmogorov's scaling invariance seems an obvious expectation. It is, 
however, refuted by experimental 
data \cite{benzi:cili:baudet:chav:95,kurien:srini:01,frisch}. 
$\zeta _p $ is not a 
linear function of $p$ (Fig. \ref{fig:one}), and the shape of the PDF varies with scale 
as in Fig. \ref{fig:two}. The deviation of $\zeta_p$ from 
$p/3$ is the anomalous 
scaling. 

Naturally, the central question is this: Why no scaling invariance? We 
suggest the simple answer that scaling invariance emerges in the PDF not for 
$\delta v$, but rather for a different function.  
This function is easiest to find in wave number space. There, 
$\mathbf{k} \cdot {\mathbf{\widehat u} }=0$ together with 
${\mathbf{\widehat u}} (-\mathbf{k} )={\mathbf{\widehat u}} 
^\ast (\mathbf{k} )$ leaves two real variables at each 
$\mathbf{k} $. Equivalently, the complex helical waves decomposition \cite{lesieur} expresses 
$\mathbf{\widehat u}(\mathbf{k})$ using a right and left-handed amplitude. 
Let $\xi _k $ be their joint PDF in wave number shell $k=\left| 
{\mathbf{k} } \right|$. Using polar form instead of two real 
amplitudes, we can expand azimuthally 
\begin{eqnarray}
\xi _k (r,\theta)=\sum_{m=0}^\infty {\xi _k^{(m)} (r)\cos (m(\theta -\theta _m } )). \nonumber
\end{eqnarray}
In shell model statistics, we have found scaling invariance of $\xi _k^{(0)} 
\circ \, \exp $, which is the mean radial profile of $\xi _k $ on a logarithmic 
abscissa. This paper shows how scaling invariance of $\xi _k^{(0)} \circ \,
\exp $ leads to a theory of anomalous scaling. We do not consider other $\xi 
_k^{(m)} $ here. 

Let us first consider the shell model data. Although crude models of the 
Navier-Stokes equations in wave number space, shell models \cite{biferale:03,bjpv,lorenz} allow us 
to gather turbulence statistics at asymptotically high Reynolds numbers. 
Moreover, shell models exhibit anomalous scaling \cite{biferale:03,bjpv}. 
We use Zimin's shell 
model \cite{zimin:huss,mela:fabi:02,mela:fabi}. Built on complex helical waves, it inherits the correct 
balance of left- and right-handedness from Navier-Stokes. Wavelets partition 
wave number space into shells $2^n\pi \le \left| {\mathbf{k} } 
\right|<2^{n+1}\pi $, and, in the spirit of Lorenz \cite{lorenz}, two collective 
variables $R_n $ and $L_n $ model all wavelet coefficients within each 
shell. Using toroidal $S_n =R_n +L_n $ and poloidal $D_n =R_n -L_n $ 
velocities, the evolution equations read
\begin{widetext}
\begin{equation}
\label{eq1}
\frac{dS_n }{dt}=2^{5n/2}\sum\limits_{m=-\infty }^\infty {T_m } \left( 
{\frac{S_n S_{n-m} }{2^{4m}}-\frac{D_n D_{n-m} }{2^{5m}}-2^{3m/2}S_{n+m}^2 
+2^{3m/2}D_{n+m}^2 } \right)-\nu 4^n2\pi ^2S_n +F_{S,n} ,
\end{equation}
\begin{equation}
\label{eq2}
\frac{dD_n }{dt}=2^{5n/2}\sum\limits_{m=-\infty }^\infty {T_m } \left( 
{\frac{S_n D_{n-m} }{2^{5m}}-\frac{D_n S_{n-m} }{2^{4m}}} \right)-\nu 
4^n2\pi ^2D_n +F_{D,n} ,
\end{equation}
\end{widetext}
where $T_m =0$ for $m<-1$, $T_{-1} =0.1935$, $T_0 =0$, and $T_m 
=2^{5/2}T_{-1} $ for $m\ge 1$; $\nu $ is the kinematic viscosity; and 
$F=(F_{S,n} ,F_{D,n} )$ is the forcing. In polar form, we have $(S_n ,D_n 
)=A_n (\cos \vartheta _n ,\sin \vartheta _n )$, where $A_n $ is the shell 
amplitude. The shell energy is $E_n =\rho _n A_n^2 /4$, and the enstrophy $Z_n 
=\pi ^24^n\rho _n A_n^2 $, where $\rho _n \equiv 7\pi 8^n/18$ is the wavelet 
density. With $E=\sum {E_n } $ and $Z=\sum {Z_n } $, we form the Reynolds 
number $R_\lambda =\left\langle E \right\rangle \nu ^{-1}(20/(3\left\langle Z 
\right\rangle ))^{1/2}$. Our focus is equilibrium statistics for isotropic 
turbulence in the classical setting of a fixed integral scale (e.g., fixed 
box size) and steady helicity-free forcing at the largest scale. Thus, we 
truncate (\ref{eq1})-(\ref{eq2}) so $n\ge 0$ and set the forcing as $F_{S,n} =-\delta _{n0} $, 
$F_{D,n} \equiv 0$. Using VODE \cite{vode}, we generate an equilibrium ensemble of 
260,000 realizations starting from random large scale initial conditions 
($R_\lambda \simeq 3\times 10^6$ and $\nu /\left| F \right|=1.3\times 
10^{-5})$.

\begin{figure}
\begin{center}
\draftfigb{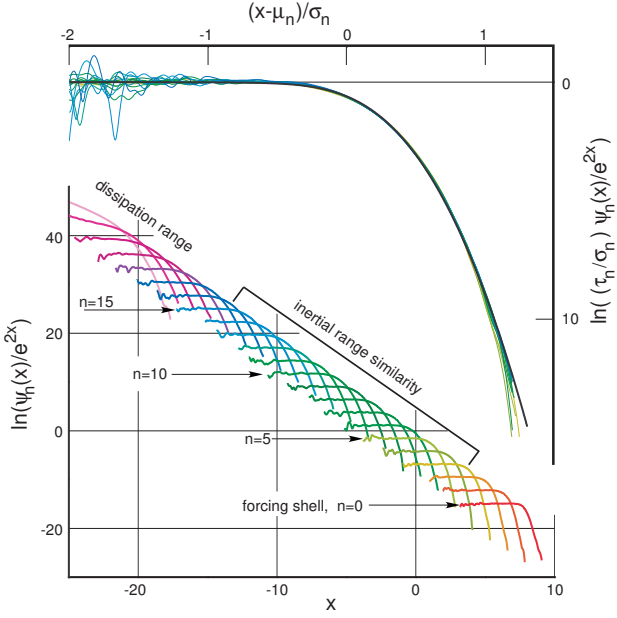}
\caption{(Color online) Statistical shell model data. The top panel shows the 
collapse in the inertial range and our theoretical curve (heavy black).
\label{fig:three}
}
\end{center}
\end{figure}

Let $\psi _n $ be the PDF of $\ln A_n $, i.e. $\psi _n (x)dx\equiv \Pr 
\{x<\ln A_n <x+dx\}$. We compute $\psi _n $ from our ensemble using kernel 
density estimation. The result (Fig. \ref{fig:three}) reveals scaling invariance in the 
inertial range: shifts and abscissa-scaling collapse the graphs of $\psi _n 
(x)/e^{2x}$ to a single curve. Thus, one function $f $ generates $\psi _n $ by the 
similarity:  
$\psi _n (x)=e^{2x}\tau _n f\left( (
x-\mu _n )/\sigma _n  \right)/\sigma _n $. This formula expresses scaling 
invariance of $\xi _k^{(0)} \circ \,\exp $. Let $\phi _n $ be the PDF for $A_n 
$. Then, $\phi _n (x)dx=\Pr \{x<A_n <x+dx\}= \quad \Pr \{\ln x<\ln A_n <\ln 
x+d\ln x\}$ so that $\phi _n (x)=\psi _n (\ln x)/x$, and our similarity 
becomes 
\begin{equation}
\label{eq3}
\frac{\phi _n (x)}{x}=\frac{\tau _n }{\sigma _n }g\left( {\left( 
{\frac{x}{\tilde {\mu }_n }} \right)^{1/\sigma_n }} \right),
\end{equation}
where $g\equiv f\circ \,\ln $ and $\tilde {\mu }_n \equiv e^{\mu _n }$. 

\begin{figure}
\begin{center}
\draftfig{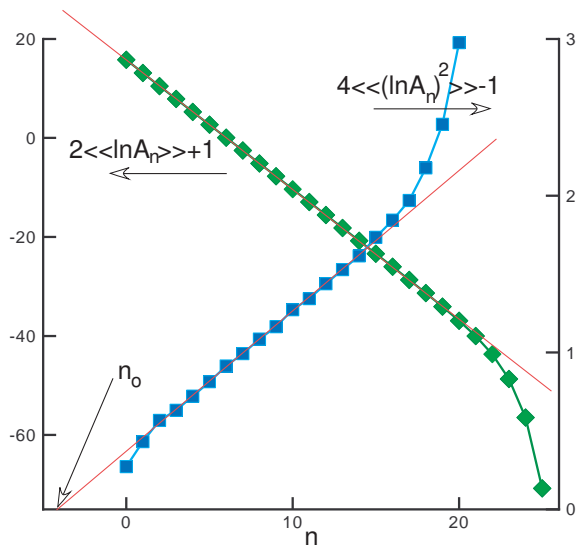}
\caption{(Color online) 
Cumulant data used to identify $n_0$, $\beta$, $C_3$, $\zeta_{-2}$ for the 
shell model.
\label{fig:four}
}
\end{center}
\end{figure}

Now we develop the theory. Motivated by the shell model, 
let us assume (\ref{eq3}) 
and seek those functions $g$ allowing 
\begin{eqnarray}
\mathcal{S}_p (n)=C_p 2^{-\zeta _p (n-n_0 )}. \nonumber
\end{eqnarray}
We 
use Mellin transforms, i.e. 
\begin{eqnarray}
G(z)\equiv \EuFrak{M}[g(x),z]=\int_0^\infty 
{x^{z-1}g(x)dx}, \nonumber
\end{eqnarray}
to express $\mathcal{S}_p (n)$ as moments of $\phi _n $. For positive 
constants $a$ and $q$, operational rules \cite{ober:74} include 
$\EuFrak{M}[g(ax),z]=a^{-z}G(z)$ 
and $\EuFrak{M}[g(x^q),z]=G(z/q)/q$. Using these, 
\begin{eqnarray}
\label{eq4}
\mathcal{S}_p (n)&=\rho _n^{p/2} \left\langle {A_n^p } \right\rangle  =\rho _n 
^{p/2}\EuFrak{M}[\phi _n (x)/x,p+2] \nonumber\\ 
 &=\rho _n ^{p/2}\tau _n \tilde {\mu }_n 
^{p+2}G((p+2)\sigma _n ).
\end{eqnarray}
Using $\mathcal{S}_0 (n)\equiv 1$, we eliminate $\tau _n $ so that $\mathcal{S}_p (n)=\rho _n 
^{p/2}\tilde {\mu }_n^p G((p+2)\sigma _n )/G(2\sigma _n )$. Setting $p=3$ 
yields $\tilde {\mu }_n =C_3^{1/3} \rho _n^{-1/2} \left( {{G(2\sigma _n )} 
\mathord{\left/ {\vphantom {{G(2\sigma _n )} {G(5\sigma _n )}}} \right. 
\kern-\nulldelimiterspace} {G(5\sigma _n )}} \right)^{1/3}2^{-\xi _3 (n-n_0 
)/3}$. Consequently, 
\begin{eqnarray}
\mathcal{S}_p (n)=C_3^{p/3} 2^{-\xi _3 p(n-n_0 )/3}\left( 
{\frac{G(2\sigma _n )}{G(5\sigma _n )}} \right)^{p/3}\frac{G((p+2)\sigma _n 
)}{G(2\sigma _n )}. \nonumber
\end{eqnarray}
The assumption $\mathcal{S}_p (n)=C_p 2^{-\zeta _p (n-n_0 )}$ 
implies 
\begin{eqnarray}
\left( {\frac{G(2\sigma _n )}{G(5\sigma _n )}} \nonumber
\right)^{p/3}\frac{G((p+2)\sigma _n )}{G(2\sigma _n )}=e^{\mathcal{A}(p)(n-n_0 
)+\mathcal{B}(p)}
\end{eqnarray}
for some $\mathcal{A}(p)$ and $\mathcal{B}(p)$. We set $n_0 $ so that $\mathcal{B}$ contains no 
multiple of $\mathcal{A}$. With $\eta \equiv \ln \circ \, G$ we obtain 
\begin{eqnarray}
\frac{p}{3}\left( 
{\eta (2\sigma _n )-\eta (5\sigma _n )} \right)+\eta \left( {(p+2)\sigma _n 
} \right)-\eta \left( {2\sigma _n } \right)\nonumber \\ =\mathcal{A}(p)(n-n_0 )+\mathcal{B}(p). \nonumber
\end{eqnarray}
Assuming the 
existence of an invertible function $s$ such that $s(n)=\sigma _n $ we have  
$w=s(n)\Leftrightarrow n=s^{-1}(w)$ and a linear functional equation for 
$\eta $:
\begin{eqnarray}
\label{eq5}
\frac{p}{3}\left( {\eta (2w)-\eta (5w)} \right)+\eta \left( {(p+2)w} 
\right)-\eta \left( {2w} \right) \nonumber \\ =\mathcal{A}(p)(s^{-1}(w)-n_0 )+\mathcal{B}(p).
\end{eqnarray}
The homogeneous solution is $\eta (w)=aw+b$. It does not affect $\mathcal{S}_p (n)$. In 
fact, $a=b=0$ means normalization, but no loss of generality. The 
non-homogeneous solution is then $\eta (w)=d_1 w^\beta +d_2 \ln w$, where 
$d_1$, $d_2$ and $\beta $ are constants. Special solutions $\eta =d_1 \left( 
{\ln w} \right)^2+d_2 \ln w$ and $\eta =d_1 w\ln w+d_2 \ln w$ substitute for 
the degenerate cases $\beta =0$ and $\beta =1$. Shell model data 
indicate $\beta \approx 1.83$, so we present analysis for $\beta \ne 0,1$. 
That is, $G(z)=e^{\eta (z)}=z^{d_2 }e^{d_1 z^\beta }$.
Experimental \cite{sief:peinke:04} and (our) computational evidence show that the joint PDF 
$\xi _k $ is finite and non-zero at the origin. Thus, we add the theoretical 
assumption of a finite non-zero value for $g(x)=\EuFrak{M}^{-1}[z^{d_2 }e^{d_1 
z^\beta },x]$ as $x\to 0^+$. This implies $d_2 =-1$. Inserting $\eta =d_1 
w^\beta -\ln w$ into (\ref{eq5}), we obtain $\mathcal{B}(p)=-p\ln (2/5)-\ln (p/2+1)$, 
$\mathcal{A}(p)=d_4 
\left( {\frac{p}{3}(2^\beta -5^\beta )+(p+2)^\beta -2^\beta } \right)$, and 
$\sigma _n =s(n)=\left( {d_4 /d_1 } \right)^{1/\beta }\left( {n-n_0 } 
\right)^{1/\beta }$, where $d_4 $ is constant. Expressing $d_4 $ in terms of 
$\zeta _{-2} $, we find for $p\geq -2$, 
\begin{equation}
\label{eq6}
\zeta _p =(\zeta _3 +\frac{3}{2}\zeta _{-2} )\left[ {\frac{(p/2+1)^\beta 
-(p/2+1)}{\left( {5/2} \right)^\beta -5/2}} \right]-\frac{p}{2}\zeta _{-2} ,
\end{equation}
\begin{equation}
\label{eq7}
C_p =C_3^{p/3} \left( {\frac{5}{2}} \right)^{p/3}\frac{2}{p+2},
\quad
p>-2.
\end{equation}
Serving only to normalize $\sigma_n $, the constant $d_1 $ is absent from 
$\zeta _p $ and $C_p $. We choose $d_1 $ so that $\left( {d_4 /d_1 } 
\right)^{1/\beta }=1/2$. With $\kappa \equiv (\zeta _3 +3\zeta _{-2} 
/2)/((5/2)^\beta -5/2)$ we have 
\begin{equation}
\label{eq8}
G(z)=z^{-1}2^{-\kappa z^\beta }.
\end{equation}
All together, we have five parameters: $C_3 ,\zeta _{-2} ,\zeta _3 ,\beta 
,n_0 .$ Kolmogorov's 4/5-law implies $\zeta _3 =1$. To find the others, we 
use cumulants of $\ln A_n $. Let $\Psi _n (s)$ be the characteristic 
function of $\psi _n (x)=e^x\phi _n (x)$. By substituting $x=\ln u$, we 
obtain $\Psi _n (s)=\EuFrak{M}\left[ {\phi _n (u),is+1} \right]=\rho _n^{-is} 
\mathcal{S}_{is} (n)=\rho _n^{-is} C_{is} 2^{-\zeta _{is} (n-n_0 )}$. Clearly, $\ln 
\Psi _n (s)$ is a linear function of $n$, as are the cumulants $\left\langle 
{\left\langle {\left( {\ln A_n } \right)^m} \right\rangle } \right\rangle 
\equiv \left[ {\left( {-id/ds} \right)^m\ln \Psi _n (s)} \right]_{s=0} $. In 
particular, $2\left\langle {\left\langle {\ln A_n } \right\rangle } 
\right\rangle +1=-\ln (7\pi /18)-3n\ln 2+2\ln (C_3 5/2)/3-(\kappa (\beta 
-1)-\zeta _{-2} )(n-n_0 )\ln 2$, and $4\langle {\langle {\left( 
{\ln A_n } \right)^2} \rangle } \rangle -1=-\kappa \beta (\beta 
-1)\left( {\ln 2} \right)\left( {n-n_0 } \right)$. We plot $\left\langle 
{\left\langle {\ln A_n } \right\rangle } \right\rangle $ and $\langle 
{\langle {\left( {\ln A_n } \right)^2} \rangle } \rangle $ 
for our shell model data in Fig. \ref{fig:four}. The data clearly identifies two lines. 
Using them, we find $C_3 =3.85$, $\zeta _{-2} =-0.825$, $\beta =1.83$, 
$n_0 =-3.7$. 
These values collapse the data in Fig. \ref{fig:three} with 
\begin{eqnarray}
&\tau _n /\sigma _n =2\rho 
_n \left( {5C_3 /2} \right)^{-2/3}2^{-(n-n_0 )\zeta _{-2} }, \nonumber \\
&\mu _n =\ln 
\left[ {\left( {5C_3 /2} \right)^{1/3}\left( {2^{(\kappa +\zeta _{-2} 
)(n-n_0 )}/\rho _n } \right)^{1/2}} \right], \nonumber \\
&\sigma _n =\frac{1}{2}(n-n_0 
)^{1/\beta }. \nonumber 
\end{eqnarray}
As the virtual origin of the inertial range scaling, $n_0$
identifies a turbulence length scale exclusively from inertial range 
properties. 

Our theory yields 
\begin{eqnarray}
\frac{\phi _n (x)}{x}= 
\frac{\tau _n }{\sigma _n }\times &\nonumber \\
\EuFrak{M}^{-1}\hspace{.2em}&\hspace{-1em}\left [  z^{-1}e^{z^\beta \text{sign}(\beta -1)},\left( 
\frac{x}{\tilde {\mu }_n } \right)^{\left( {\left| \kappa \right|\ln 2} 
\right)^{-1/\beta }/\sigma _n } \right ], 
\end{eqnarray}
where numerical techniques \cite{fabi:mela:05} provide 
the inverse Mellin transform. Recall that $\phi _n (x)/x$ represents the 
first term $\xi _k^{(0)} $ in the azimuthal expansion of the joint PDF $\xi 
_k $, so to compare with one-dimensional PDFs from the literature, 
e.g.  \cite{frisch,biferale:03,bjpv}, we integrate $\phi _n \left( {\left( {S_n^2 +D_n^2 } 
\right)^{1/2}} \right)$ over $D_n $. The result is an approximation to the 
PDF for $S_n $ (Fig. \ref{fig:two}). The shape of the PDF differs between shells. 
It is roughly Gaussian for small $n-n_0 $, but develops a sharp peak and 
broad tails as $n$ increases. These are classical characteristics of anomalous 
scaling. 

Theory often requires that $\zeta _p $ increases monotonically \cite{frisch}. This is a 
consequence of assuming an upper bound for the velocity so as to avoid 
supersonic speeds \cite{frisch}. Correspondingly, the PDF for the velocity increment can 
not have an infinite tail, but must have compact support. Our function $g$ has 
compact support when $\beta <1$, but not when $\beta >1$. Importantly, our 
data yields $\beta \simeq 1.83>1$. Consequently, $\zeta _p $
 has a maximum and 
even decreases to negative values when $p$ is large. Negative 
values of $\zeta _p $ are not unphysical. In fact, they occur in 2D 
turbulence \cite{eyink:96}. As a mathematical problem, turbulence described by the 
incompressible Navier-Stokes equations does not know of supersonic speeds 
and has its own statistics. We can not impose constraints on that 
statistics. It is a different matter that the incompressible equations 
incorrectly describe fluid flows at supersonic speeds. It is wrong to 
require $d\zeta _p /dp>0$ simply because the condition imposes compact support on the PDF. In 
contrast, $d^2\zeta _p /dp^2<0$ follows from $\left\langle {\left\langle 
{A_n^2 } \right\rangle } \right\rangle >0.$

Our theory has scaling invariance in the form of the 
similarity formula (\ref{eq3}), 
but also reproduces the well known anomalous scaling when expressed in 
traditional variables (Fig. \ref{fig:two}). 
Thus, by choosing the right variable, 
anomalous scaling disappears and we obtain a collapse across scales as in 
Figure 3. Our theory rests on four inertial range assumptions: scaling 
invariance of $\xi _k^{(0)} \circ\, \exp$ (Eqn.(\ref{eq3})), existence of the 
exponents $\zeta _p $, a finite $\xi _k^{(0)} (0)\ne 0$, and an invertible 
$\sigma _n $. These assumptions hold for our shell model. We can only 
conjecture that they also hold for Navier-Stokes. 
To test our theory, one should aim to reproduce the collapse in 
Fig.~\ref{fig:three}. That 
is, sample $\ln \left( {E\left(| \mathbf{k} |\right)} \right)$ 
for various $k=|\mathbf{k}|$ in the 
inertial range, $E\left( |\mathbf{k}| \right)$ being the 
three-dimensional energy. Call 
the PDF $\psi _k $ and plot the lower part of Fig.~\ref{fig:three} 
with $\psi _k $ replacing $\psi _n $. If curves for different $k$ 
can be brought to coalesce 
by shifts and horizontal scaling, then our theory applies and gives the 
analytical expression for $\psi _k $ as well as many other formulas.

\bibliographystyle{apsrev}

\end{document}